\documentclass[]{spie}

\usepackage{amsmath,graphicx,tabularx,ulem,multirow}

\title{Practical Coherent Integration with the NPOI}

\author{Anders M. Jorgensen\supit{a}, Dave Mozurkewich\supit{b},
  Henrique Schmitt\supit{c,d}, Robert Hindsley\supit{c},\\ J. Thomas Armstrong\supit{c}, Thomas A.
  Pauls\supit{c}, D. Hutter\supit{e}
\skiplinehalf
  \supit{a}New Mexico Institute of Mining and Technology, Socorro, NM, USA\\
  \supit{b}Seabrook Engineering, Seabrook, MD, USA\\
  \supit{c}Naval Research Laboratory, Washington, DC, USA\\
  \supit{d}Interferometrics, Inc., Herndon, VA, USA\\  
  \supit{e}Naval Observatory Flagstaff Station, Flagstaff, AZ, USA\\
  \footnotemark[0]}

\begin{document}

\twocolumn[{\csname @twocolumnfalse\endcsname
\vspace{0.2in}

\maketitle

\begin{abstract}
In this paper we will discuss the current status of coherent
integration with the Navy Prototype Optical Interferometer
(NPOI)\cite{armstrong:1998}. Coherent integration relies on being able
to phase reference interferometric measurements, which in turn relies
on making measurements at multiple wavelengths. We first discuss the
generalized group-delay approach, then the meaning of the resulting
complex visibilities and then demonstrate how coherent integration can
be used to perform very precision measurement of stellar
properties. For example, we demonstrate how we can measure the
diameter of a star to a precision of one part in 350, and measure
properties of binary stars. The complex phase is particularly
attractive as a data product because it is not biased in the same way
as visibility amplitudes.
\end{abstract}

}]

\section{Introduction}

At the 2004 and 2006 SPIE meetings we demonstrated how to coherently
integrate NPOI\cite{armstrong:1998} data by fitting a model of the
fringes to the raw data and using the fitting parameters to coherently
integrate the complex visibilities from individual frames
\cite{jorgensen:2004, jorgensen:2006}. This discussion was also
published in a separate paper \cite{jorgensen:2007}. In this paper we
take a different approach to fringe tracking which is an extension of
the conventional group-delay approach, but including the atmospheric
dispersion as well as time-dependence in the group-delay
calculation. This is discussed in
Section~\ref{section_group_delay}. We then discuss the meaning of the
resulting coherently integrated complex visibilities, including the
phase (Section~\ref{section_coherent_phase}), and the amplitude
(Section~\ref{section_coherent_amplitude}). We then show two examples
of how to use complex visibilities for scientific measurements,
including an precise diameter measurement
(Section~\ref{section_diameter}) and precise measurements on a binary
star (Section~\ref{section_binary}), before concluding.

\begin{figure}
\includegraphics[width=\linewidth]{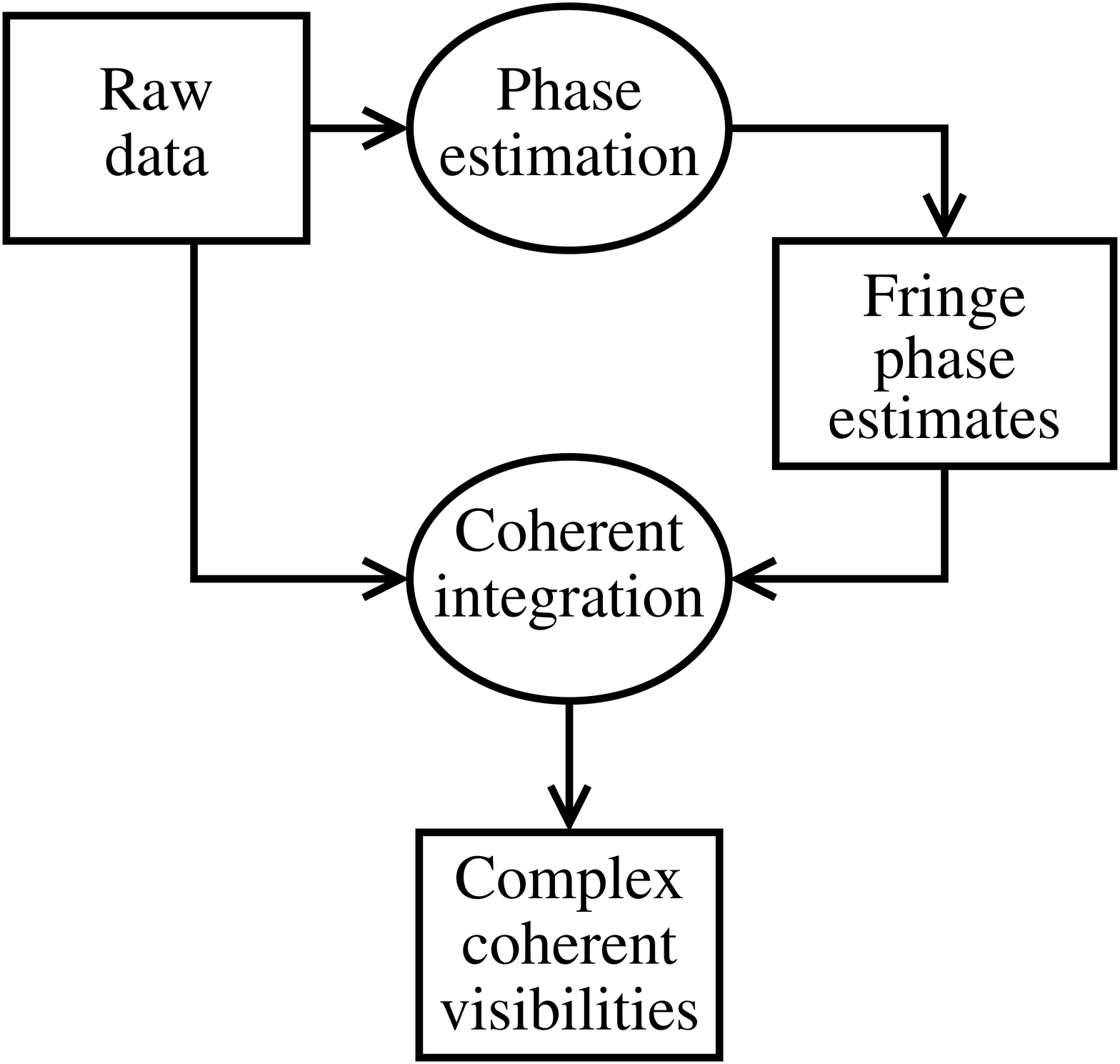}
\caption{\label{figure_ci_overview} An overview of the coherent
  integration procedure. It consists of two primary steps. The first
  is the determination of the fringe phase, and the second is the
  coherent integration using the measured fringe phase.}
\end{figure}

The Navy Prototype Optical Interferometer (NPOI)
\cite{armstrong:1998}, located in Flagstaff, AZ, measures fringes by
scanning the optical path-difference across several fringe periods
(usually between one and eight) and using photon-counting avalanche
photo diodes to detect the photons and bin them according to fringe
phase at up to 32 different wavelengths in three different
spectrographs. This produces a regular array every 2 ms with fringe
phase along one dimension and wavelength along the other dimension,
making it easy to process the arrays with Fourier-based techniques.
For technical reasons the NPOI data has been restricted to 16
wavelength channels in two spectrographs in recent years, and those
are the data we will show here. It is important however that NPOI
measures many channels (which we will define as approximately ten or
more) because it is this multi-wavelength capability which makes
coherent integration possible and interesting. Coherent integration
consists of two primary steps, as outline in
Figure~\ref{figure_ci_overview}. The first step uses the data to
estimate the fringe phase whereas the second step uses the estimate of
the fringe phase together with the data to produce coherently
integrated visibilities, as outline in the next two sections.

\section{Generalized group-delay fringe finding}
\label{section_group_delay}

The traditional group-delay is defined as the vacuum delay, $d$, which
best aligns the fringes in different wavebands. If a delay is applied,
a fringe at wavelength $\lambda$ is rotated by

\begin{equation}
  \label{eq_vacuum_delay}
  \theta=\frac{2\pi d}{\lambda}
\end{equation}

\noindent We can add the visibilities at different wavelengths,

\begin{equation}
  X+iY=\sum_i (X+iY)_i e^{-\theta_i}
  \label{eq_simple_gd_sum}
\end{equation}

\noindent where $X$ and $Y$ are the cosine and sine transforms of the
photon data, and related to the visibility as
$\tilde{V}=\frac{X+iY}{N}$, where $N$ is the total number of
photons. The sum is over different wavelength channels. The
group-delay is then defined as the value of $d$ which maximizes
$\left|X+iY\right|^2$. An algorithm similar to this is implemented in
the NPOI real-time fringe-tracking system \cite{benson:1998}, and the
best group-delay can be found quickly because it involves a linear
search over a single parameter.

We can generalize this simple approach to fringe tracking by including
more parameters in Equation~\ref{eq_vacuum_delay}, allowing them to
vary as a function of time, and making the sum in
Equation~\ref{eq_simple_gd_sum} to be over multiple consecutive
exposures.

First, a more realistic functional form for the variation of the phase
with wavelength should include the dispersive atmosphere,

\begin{equation}
  \label{eq_vac_plus_atm}
  \theta_i=\frac{2\pi(d+(n(\lambda_i)-1)a)}{\lambda_i}
\end{equation}

\noindent We can also include time-variation of the parameters $d$,
and $a$. We allow them to vary as Legendre polynomials (which are
defined in the interval $[-1;1]$), e.g.

\begin{equation*}
  d(t-t_0)=\sum_{i=0}^{n_d-1} e_iL\left(\frac{2(t-t_0)}{T}\right)
\end{equation*}

\noindent and

\begin{equation*}
  a(t-t_0)=\sum_{i=0}^{n_a-1} b_iL\left(\frac{2(t-t_0)}{T}\right)
\end{equation*}

\noindent Where we are modeling the time-variation of $d$ and $a$ over
the time-interval of length $T$. We can then determine a set of
parameters, $\{e_i\}$ and $\{b_i\}$ which best models the wavelength
and time-variation of the fringes over a short time-interval. 

We find this best set of model parameters by maximizing the quantity
$\left|X+iY\right|^2$, where

\begin{equation*}
  X+iY=\sum_{j=1}^M\sum_{i=1}^N \left(X_{ij}+iY_{ij}\right)e^{-\theta_{ij}}
\end{equation*}

\noindent Where $j$ counts time, and $i$ counts wavelengths. Our claim
is then that fitting this more complicated fringe phase function to a
set of frames can produce a better estimate of the phase of the
central frame of the set. We estimate $d_{ij}$ and $a_{ij}$ for the
central frame. In this way we estimate $d$ and $a$ for nearly all the
data frames in the data sets. We also record the phase
$\phi=\angle X+iY$.

\section{Coherent integration}

To coherently integrate we rotate the individual complex quantities
$(X_{ij}+iY_{ij})$ by the angle

\begin{equation}
  \theta_{ij}=\frac{2\pi\left(d_{j}+a_{j}(n(\lambda_i)-1)\right)}{\lambda_i}-\phi_{j}
\end{equation}

\noindent where $i$ counts wavelength, and $j$ counts the frames
(exposures) over which we are coherently integrating (typically 10's
of thousands). The coherent integration looks like this

\begin{equation}
  X_i+iY_i=\sum_{j=1}^M \left(X_{ij}+Y_{ij}\right)e^{\theta_{ij}}
\end{equation}

\noindent We also sum the photon counts,

\begin{equation*}
  N_i=\sum_{j=1}^M N_{ij}
\end{equation*}

\noindent The complex visibility at each wavelength can then be
computed as

\begin{equation*}
  \tilde{V_i}=\frac{X_i+iY_i}{N_i}
\end{equation*}

\noindent and the squared visibility and visibility phase (at each
wavelength) can be computed as

\begin{equation*}
  \label{eq_coherent_v2}
  V_i^2=4\frac{X_i^2+Y_i^2-N_i}{N_i^2}
\end{equation*}

\noindent (assuming Poisson statistics) and

\begin{equation*}
  \phi_i=\tan^{-1}(Y_i/X_i)
\end{equation*}

\noindent where one must be careful to consider the quadrant of
$\phi_i$ for it to range over the full $[0;2\pi[$ range, for example
    by using the function \texttt{atan2(Y,X)} available in some form
    in most computer languages. To consider non-Poisson statistics one
    must replace the subtracted $N$ in Equation~\ref{eq_coherent_v2}
    with the actual estimated Bias. However, in most practical
    situations, the bias is smaller than the uncertainty and can thus
    be ignored. For that reason it is usually not necessary to
    consider the details of the bias statistics either. Effectively,
    coherently integrated squared visibilities do not have a bias - at
    least not one large enough to need to correct, except in cases of
    extremely small visibility amplitudes.

\begin{figure}
\includegraphics[width=\linewidth]{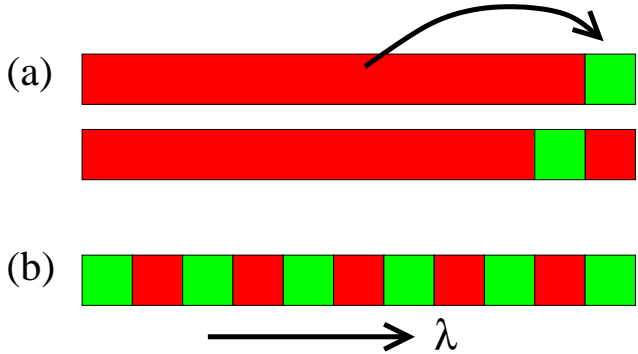}
\caption{\label{figure_w_bootstrap} Illustration of two approaches to
  fringe finding and coherent integration which bootstraps the phase in one wavelength channel from phase measurements in other wavelength channels.}
\end{figure}

When coherently integrating it is not possible to coherently integrate
on the same data used for fringe tracking. If that is done, a bias
similar to the bias in incoherent $V^2$ averaging is introduced. We
can overcome this problem by sub-dividing the data sets. For example,
we can separate the data set into even-numbered channels and
odd-numbered channels (as illustrated in
Figure~\ref{figure_w_bootstrap}b), track fringes in each set, and use
that information to coherently integrate the channels in the other
set. This is a form of wavelength bootstrapping. However it is not
very efficient in that for each channel we now only track fringe using
half of the available data. A more computationally intensive approach
which produces better signal-to-noise ratio is illustrated in
Figure~\ref{figure_w_bootstrap}a, the fringe finding is repeated for
each channel, each time leaving out only that channel. 

\begin{figure}
\includegraphics[width=\linewidth]{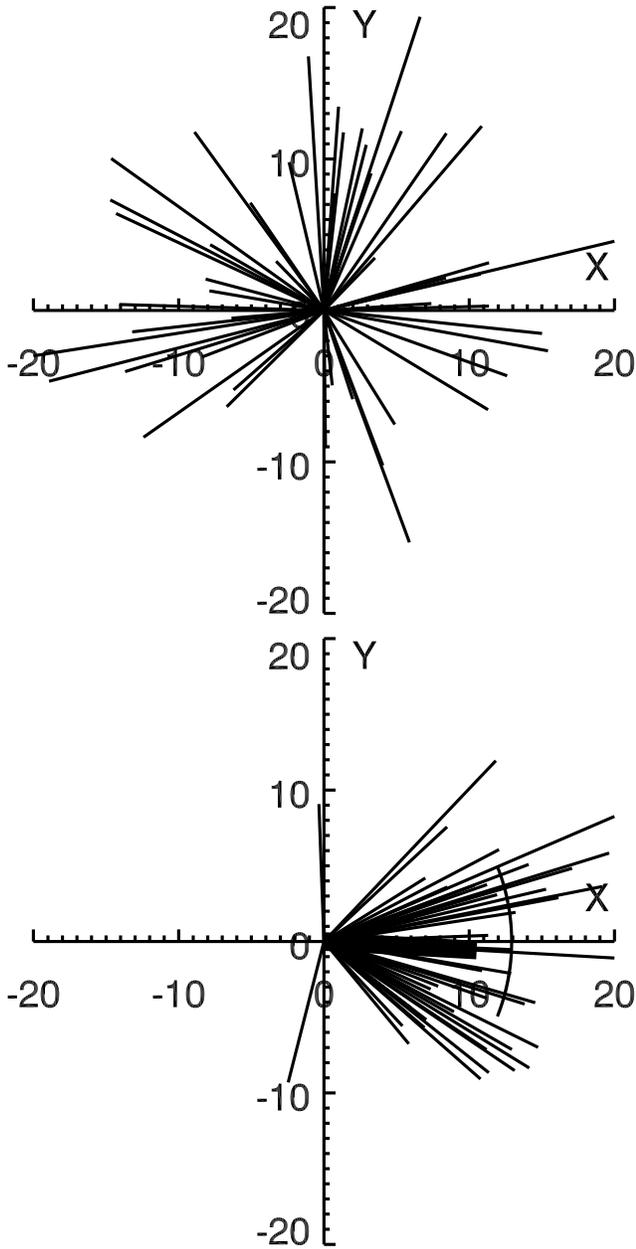}
\caption{\label{figure_phasor_rotation} Illustration of the process of
  phasor rotation and coherent integration. In the top panel are the
  raw coherent visibilities for 100 short observations of the star
  $\nu\,\text{Oph}$. In the bottom panel the visibilities have been
  rotated, and the thick phasor is their average. The circle segment
  shows the average amplitude of the individual visibilities,
  demonstrating that phase noise reduces the coherently integrated
  visibility.}
\end{figure}

An example of what the coherent integration process looks like is
shown in Figure~\ref{figure_phasor_rotation}. This figure illustrates
the rotation and coherent averaging of complex visibilities. In the
top panel are plotted the raw coherent visibilities. In the bottom
panel are plotted the same visibilities after rotation as well (thick
vector) the average of those visibilities. The amplitude of the
coherent average is smaller than the average of the amplitudes of the
individual complex visibilities, with the difference due to the
imperfect rotation of the complex visibilities. This effect of
reducing the visibility amplitude when coherently integrating, because
of phase noise, will be discussed in
Section~\ref{section_coherent_amplitude}.

\section{The coherent phase}
\label{section_coherent_phase}

\begin{figure}
\includegraphics[width=\linewidth]{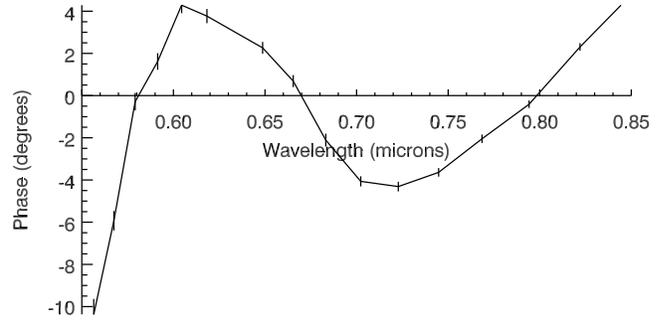}
\caption{\label{figure_instrumental_phase} Example of an instrumental
  phase measured by observing a calibrator star.}
\end{figure}

\noindent Next we consider the phase of the coherently integrated
visibilities. Following the simple procedure above, the phase on each
baseline is composed of the following components,

\begin{equation}
\label{eq_coherent_phase_components}
\phi=\phi_\text{source}+\phi_\text{instr}+\phi_\text{atm}
\end{equation}

\noindent which are the source phase (the one we are interested in),
an instrumental phase caused by mismatch between the two light paths
of the interferometer (in particular glass windows), and a residual
atmospheric phase. Note that these three phases are additive. Thus, we
can determine the instrumental phase plus an arbitrary amount of
atmospheric phase by observing a calibrator which has zero source
phase. Subtracting the measured phase of a calibrator star thus
eliminates the instrumental term, but leaves the atmospheric term. We
know the exact form of the atmospheric term, because it looks like
Equation~\ref{eq_vac_plus_atm}. Figure~\ref{figure_instrumental_phase}
shows the instrumental phase recorded from a calibrator star. An
important observation is that the phase of the coherently integrated
visibilities is not biased like the amplitude is. This means that the
raw photon-counting-based uncertainty of the phase is the actual
uncertainty. The uncertainty of the phase can be shown to be 

\begin{equation*}
  \sigma_{\phi_i}=\frac{1}{\sqrt{2N_iV_i^2}}
\end{equation*}

\noindent Where $N_i$ is the total number of photons in channel $i$
over the coherently integrated interval, and $V_i$ is the raw
coherently integrated visibility amplitude.

\section{The coherent amplitude}
\label{section_coherent_amplitude}

\begin{figure}
  \includegraphics[width=\linewidth]{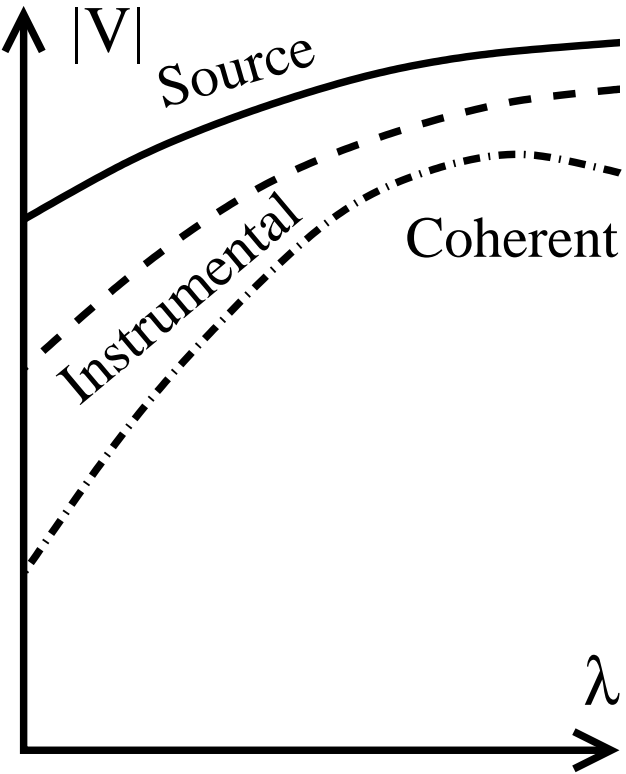}
\caption{\label{figure_ampl_reduction} Illustration of the reduction
  of the coherently integrated visibility amplitude relative to the
  raw visibility amplitude. This reduction is due to the finite
  precision in the determination of rotation phases.}
\end{figure}

\noindent The coherent amplitude is reduced due to phase noise in the
determination of the rotation phase. As illustrated in
Figure~\ref{figure_ampl_reduction} coherent integration causes an
additional amplitude reduction due to the phase noise in determining
the appropriate rotation of the individual phasors.

In order to make use of the coherently integrated visibility
amplitudes it is necessary to calibrate them and correct for the
amplitude-reducing effects of the phase noise. Once this is done, the
resulting instrumental visibilities can be calibrated in the usual
manner by dividing by the amplitude of a calibrator star. The simplest
description of the effect of the phase noise can be based on the
distribution of phase errors. If the phase errors, $\Delta\theta$ are
distributed according to $f(\Delta\theta)$, then the visibility will
be reduced by the factor $\gamma$,

\begin{equation*}
  \gamma=\int f(\Delta\theta) e^{i\Delta\theta}d\Delta\theta
\end{equation*}

\noindent In the case where $f(\Delta\theta)$ is a Gaussian with
standard deviation $\sigma_\theta$, the integral reduces to 

\begin{equation*}
  \label{eq_gaussian_phase_noise}
  \gamma=e^{-\sigma_\theta^2}
\end{equation*}

\begin{figure}
\includegraphics[width=\linewidth]{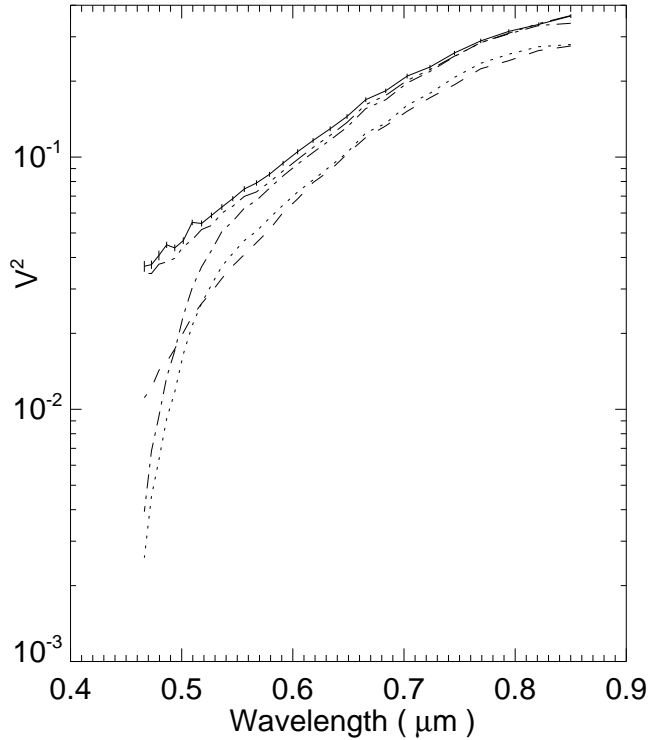}
\caption{\label{figure_amplitude_self_cal} Amplitude calibration of
  coherently integrated squared visibilities. The solid curve is the
  true instrumental squared visibility. The dotted curve is the
  coherently integrated squared visibility using a vacuum variation of
  phase only (Equation~\ref{eq_vacuum_delay}), whereas the dashed
  curve uses a model which contains both vacuum and atmosphere
  (Equation~\ref{eq_vac_plus_atm}). The dot-dashed curve is the
  correction of the dotted curve, whereas the triple-dot-dashed curve
  is the correction of the dashed curve.}
\end{figure}

If we then assume that a Gaussian is a reasonable model for the phase
noise distribution the next question becomes how we measure
$\sigma_\theta$. Because we measure more than one (at least two)
estimate for the fringe phase, as illustrated in
Figure~\ref{figure_w_bootstrap}, we can also estimate the phase noise
as the difference between these multiple estimates. However, when
doing this it is important that the function used to represent the
fringe phase is a true description of the fringe phase with the
appropriate number of degrees of freedom, for example the expression
in Equation~\ref{eq_vac_plus_atm} which contains both a vacuum path
and an atmosphere path. Figure~\ref{figure_amplitude_self_cal}
illustrates how well amplitude calibration can be performed, and the
importance of choosing an appropriate model for the fringe phase as a
function of wavelength.

In Figure~\ref{figure_amplitude_self_cal} the solid curve is the true
instrumental squared visibility, which is the target for the
calibration of the coherently integrated squared visibilities. The
dotted curve represents the coherently integrated squared visibility
obtained by finding fringes with a model which tracks only vacuum
delay (Equation~\ref{eq_vacuum_delay}). The dashed curve is the
squared visibility coherently integrated when the fringe phase was
found using a model which includes both vacuum and atmospheric delay
(Equation~\ref{eq_vac_plus_atm}). We notice firs that the simpler
model performs marginally better at longer wavelengths, whereas the
more complex model performs much better at shorter
wavelengths. Because we subdivided the wavelength channels into
groups, as illustrated in Figure~\ref{figure_w_bootstrap}, we can
estimate the uncertainty with which we have estimated the phase as the
difference between the estimates of these different groups, and use
that with Equation~\ref{eq_gaussian_phase_noise} to apply a
correction. 

The dot-dashed curve is the corrected squared visibility for the case
where a pure vacuum model was use for tracking the fringes. We can see
that at the red end of the spectrum it is close to the true
instrumental visibility whereas at the blue end of the spectrum the
correction does not work. The reason for this is that the simple
vacuum-only model is not flexible enough to reproduce the motion of
the fringes at the blue end of the spectrum due to the variation of
the atmosphere in addition to the vacuum.

On the other hand, when we use a model which incorporates both a
vacuum term and a atmosphere term, and correct the squared visibility
we obtain the triple-dot-dashed curve in
Figure~\ref{figure_amplitude_self_cal}, which closely matches the true
instrumental visibility at all wavelengths. This illustrates the
importance of using a model which is able to correctly represent the
variability of the fringe phase, if that variability is to be used for
amplitude calibration.

\begin{figure*}
\includegraphics[width=\linewidth]{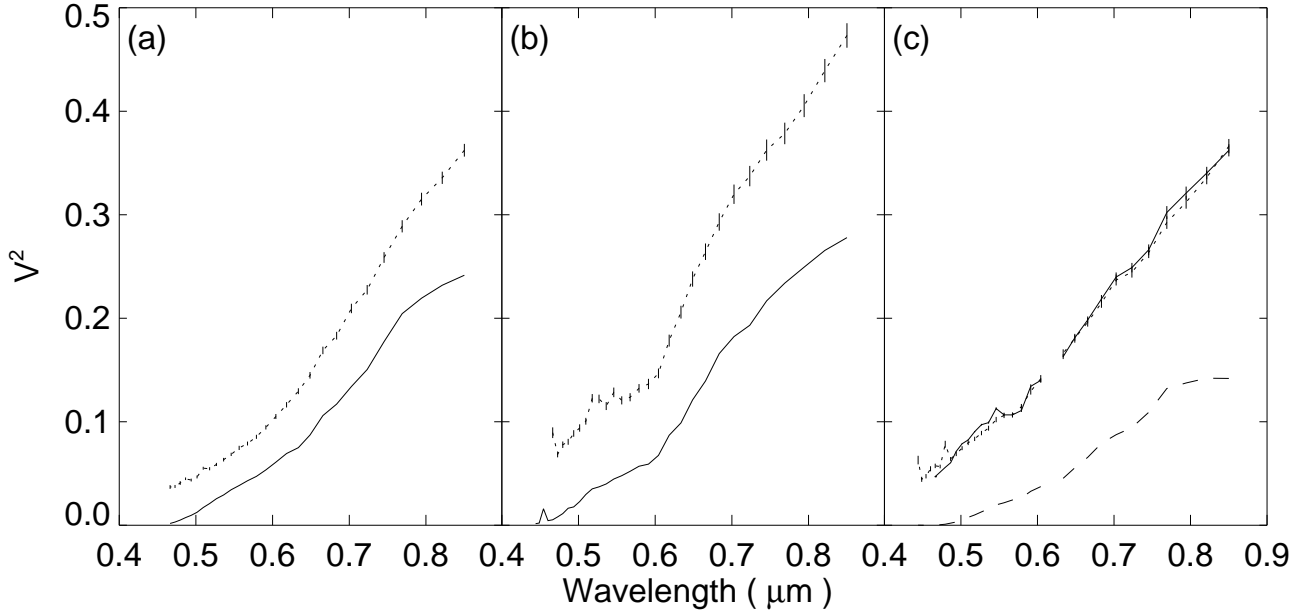}
\caption{\label{figure_amplitude_boot_cal} It is possible to calibrate
  the amplitude of a bootstrapped baselines by adding the phase noise
  from the two bootstrapping baselines in quadrature. The phase noise
  on those two baselines is in turn obtained from the ratio of the
  coherently average and incoherent squared visibilities.}
\end{figure*}

There is a case in which we are much less sensitive to the accuracy of
the model, and that is when we are attempting to calibrate a
bootstrapped baseline. One of the most powerful uses of coherent
integration is its use to improve SNR on baselines which are too faint
to have any distinguishable fringes in individual exposures because
the visibility is very small. In those cases the fringes cannot be
tracked directly, but are instead bootstrapped from two baselines with
which the baseline forms a closure triangle. If then the incoherent
visibilities as well as the coherent visibilities are available on the
two bootstrapping baselines, the ratio can be used as a measure of the
phase noise on the bootstrapping baselines. When bootstrapping the
phase noise adds in quadrature, and can then be applied to correct the
amplitude on the bootstrapped baseline. This approach is less
sensitive to using the correct function form of fringe phase versus
wavelength for tracking fringes. However, it requires the use of the
incoherent squared visibilities with associated complications and SNR
limitations. The approach is illustrated in
Figure~\ref{figure_amplitude_boot_cal}.

In Figure~\ref{figure_amplitude_boot_cal} we plot squared visibilities
as a function of wavelength for three baselines which form a closure
triangle. In this case we track the fringes on the first two baselines
(panels (a) and (b)), and use that information to coherently integrate
on the third baseline (panel (c)). In panels (a) and (b) the solid
curves are the coherently integrated squared visibilities. In
addition, we estimate the phase noise on the first two baselines by
making use of the incoherent squared visibility (dotted curves in
panels (a) and (b)), and use that information to bootstrap the phase
noise on the third baseline. We show that this is done correctly by
demonstrating that when the bootstrapped coherently integrated squared
visibility on the third baseline (dashed curve in panel (c)) is
corrected (solid curve), it agrees with the incoherent squared
visibility (dotted curve). There are thus multiple approaches to
calibrating the coherently integrated visibility amplitudes, each of
which make use of information about the distribution of phases noise in
different ways.

\section{Real-time or post-processing coherent integration?}

\begin{table*}
\centerline{\begin{tabularx}{.7\textheight}{|l|X|X|}
\cline{2-3}
\multicolumn{1}{c}{ }& \multicolumn{1}{|c|}{Positives} & \multicolumn{1}{c|}{Negatives}\\
\hline
\multirow{4}{1in}{Real-time} & \multirow{4}{\hsize}{\textbullet Single detector read} & \textbullet Fast hardware\\
& & \textbullet Prediction\\
& & \textbullet Phase-locking\\
& & \textbullet No reprocessing\\
\hline
\multirow{3}{1in}{Post-processing} & \textbullet Slow hardware & \multirow{4}{\hsize}{\textbullet Many detector reads}\\
& \textbullet Interpolation & \\
& \textbullet Envelope locking & \\
& \textbullet Reprocessing possible & \\
\hline
\multirow{3}{1in}{Both} & \multirow{3}{\hsize}{\textbullet Better SNR\\\textbullet Phase on all baselines} & \textbullet New data product\\
& & \textbullet Meaning of phase\\
& & \textbullet Effect of phase noise\\
\hline
\end{tabularx}}
\caption{\label{table_real_time_post} Comparison of the advantages and
  disadvantages of real-time and post-processing coherent
  integration. The general conclusion is that real-time coherent
  integration is likely more advantageous at infrared wavelenghts,
  whereas post-processing coherent integration is likely more
  advantageous at visible wavelengths.}
\end{table*}

\noindent In recent years there has been significant effort towards
real-time coherent integration, or long-term integration on a detector
while stabilizing fringes by some external means. Several such systems
are now coming online at the VLTI for example (See e.g. Le
Bouquin~\cite{lebouquin:2008}. This begs the question whether the NPOI
coherent integration approach of measuring many short exposures and
combining them after the fact in post-processing is a valuable
approach for the long-term or whether it is simply an intermediate
step on the way to more sophisticated hardware which performs coherent
integration in real time.

We argue that post-processing coherent integration is not an
intermediate step, but that it is in-fact an optimal approach for some
types of measurements. In that case real-time coherent integration is
also an optimal approach under certain conditions. A significant
advantage of real-time coherent integration is that it involves very
few detector reads, which means that read-noise is reduced. This
approach should therefore be selected when possible if the detectors
have high read noise, such as is often the case in the infrared bands.

On the other hand, real-time coherent integration has the disadvantage
that the fringe phase must be constantly predicted from measurements
that have already been recorded. If a complete data set of short
exposures is already available, then the fringe phase can be
interpolated from future as well as past measurements (and the present
measurements can be used as well of course), which may be a
significant advantage when fringe motion is rapid, such as is often
the case at shorter wavelengths. It is also often the case at shorter
wavelengths, that the read noise of the detectors is lower, such that
performing many detector reads does not seriously degrade the
data. Table~\ref{table_real_time_post} lists some of the advantages
and disadvantages of each approach. The conclusion is that real-time
coherent integration is likely to be more advantageous at infrared
wavelenghts, whereas post-processing coherent integration is likely to
be more advantageous at visible wavelengths.

\section{Measuring the diameter of a star}
\label{section_diameter}

\begin{table}
\centerline{\begin{tabular}{lrr}
\hline
Date & Time & Hour angle\\
\hline
2005/6/29 & 06:32 & -0.396\\
2005/6/29 & 06:51 & -0.080\\
2005/6/29 & 07:03 & 0.134\\
2005/6/29 & 07:17 & 0.347\\
2005/6/29 & 07:28 & 0.540\\
2005/7/9 & 06:12 & -0.070\\
2005/7/9 & 06:41 & 0.400\\
2005/7/9 & 06:58 & 0.680\\
2005/7/9 & 07:14 & 0.950\\
\hline
\end{tabular}}
\caption{\label{table_nu_oph} List of observations of $\nu\,\text{Ophiuchus}$.}
\end{table}

\begin{figure}
\includegraphics[width=\linewidth]{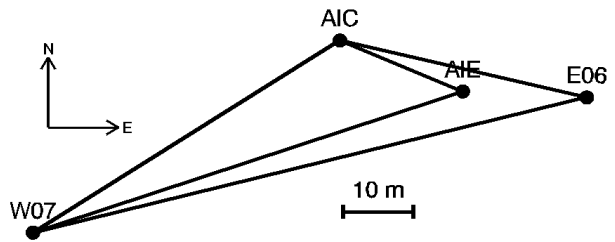}
\caption{\label{nuoph_baselines} Layout of the baselines for
  $\nu\,\text{Ophiuchus}$. The longest baseline, E06-W07, contains the
  zero-crossing.}
\end{figure}

\begin{figure*}
\includegraphics[width=\linewidth]{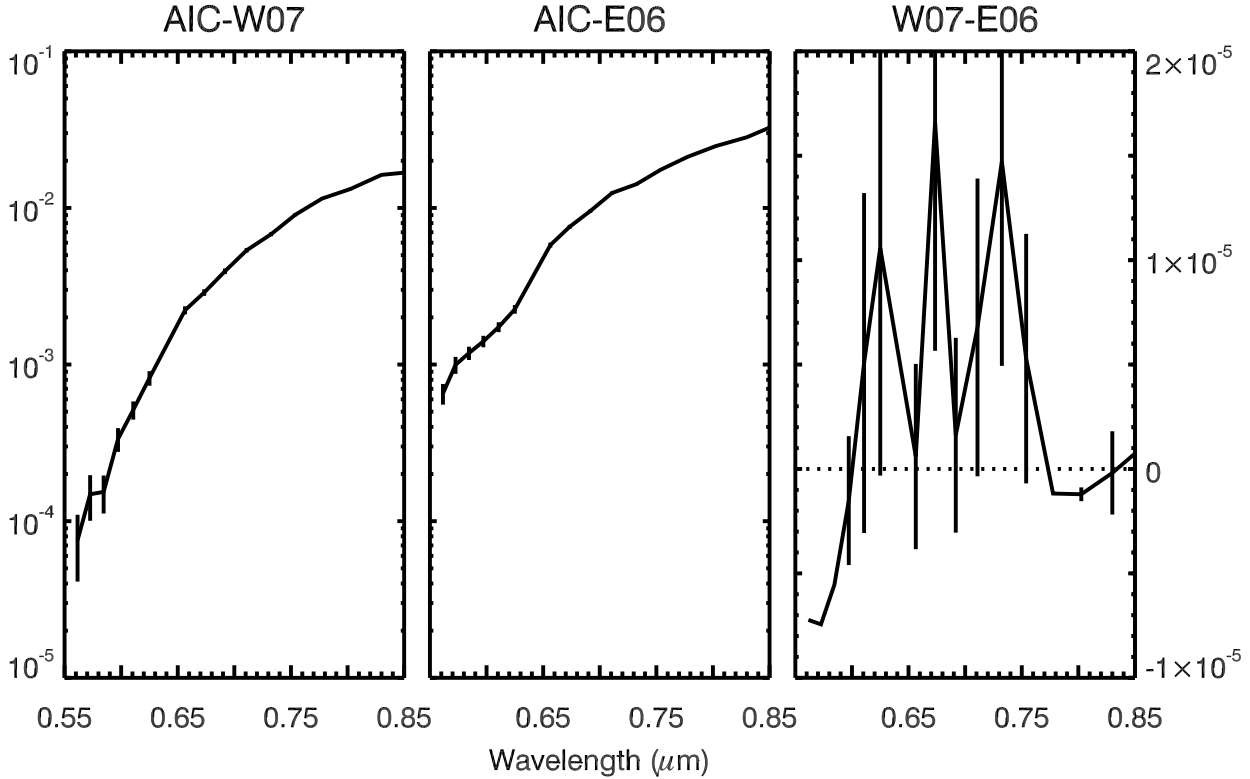}
\caption{\label{figure_baselineboot1} Coherent integration of the
  three baselines in the closure triangle separately. It can be seen
  that the SNR on the W07-E06 baseline is extremely poor.}
\end{figure*}

\begin{figure*}
\includegraphics[width=\linewidth]{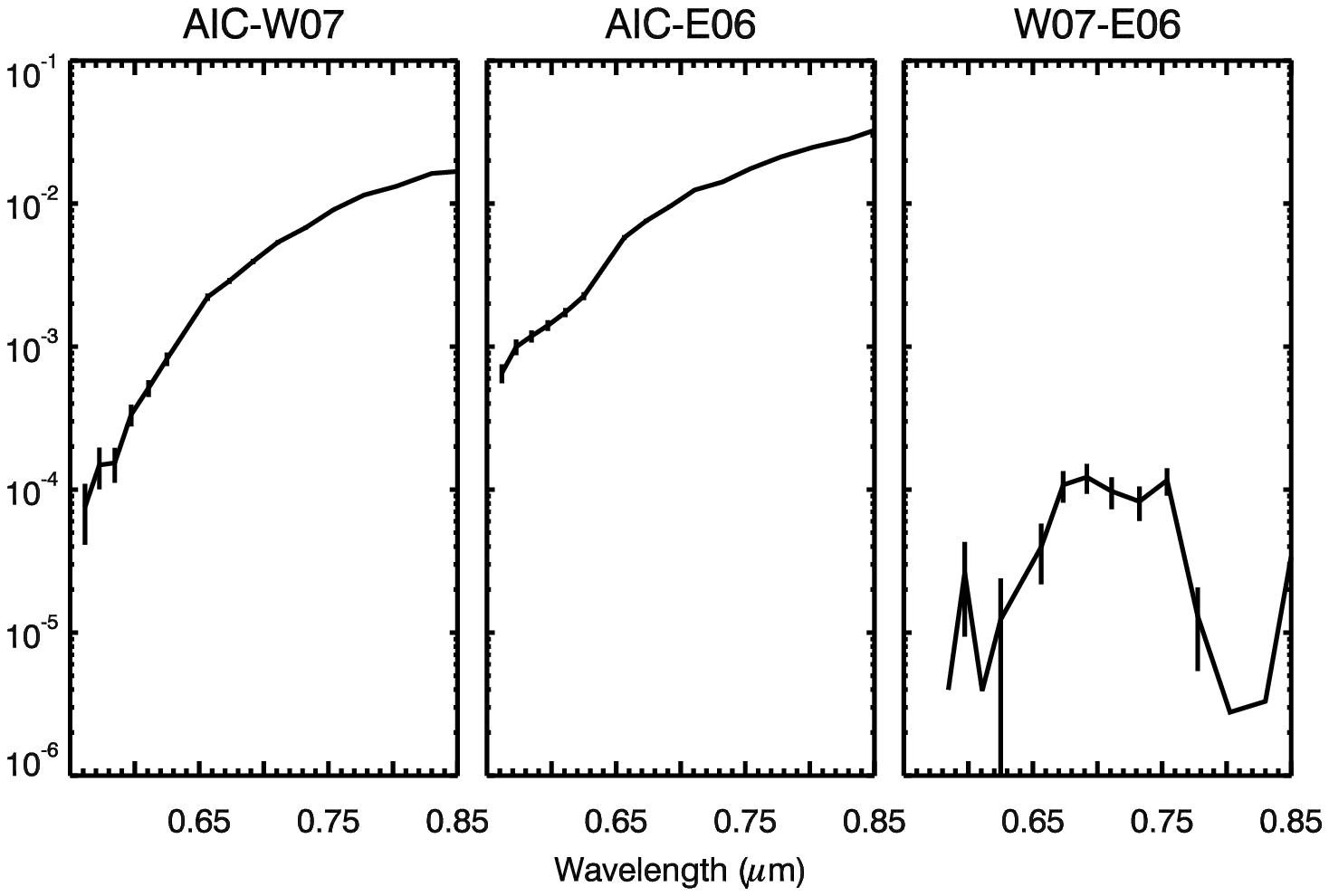}
\caption{\label{figure_baselineboot2} Coherent integration of the
  three baselines in the closure triangle. In this case the phase on
  the long baseline, W07-E06, has been bootstrapped from the two
  shorter baselines, with the result that the SNR on the baseline is
  greatly improved.}
\end{figure*}

\noindent We will illustrate the use of coherently integrated
visibilities by measuring the diameter of a binary star. We make use
of measurements of the binary star $\nu\,\text{Ophiuchus}$ on two
different days in 2005 and demonstrate how minimally calibrated data
can be used to make very high-precision
measurements. Table~\ref{table_nu_oph} lists the observations. Each
observation was performed on five baselines as illustrated in
Figure~\ref{nuoph_baselines}. The longest baseline, E06-W07, contains
a visibility zero-crossing. This zero-crossing is directly related to
the uniform-disk diameter of the star. If we can measure the
zero-crossing very accurately we can therefore measure the diameter of
the star very precisely. However, we cannot coherently integrate on
that baseline directly because the visibility is small and the SNR
thus small also. Figure~\ref{figure_baselineboot1} shows the real
component of the coherently integrated visibility in that
case. Instead we must bootstrap the phase on the long baseline from
the measured phase on the two shorter baselines with which it forms a
closure triangle. The result is shown in
Figure~\ref{figure_baselineboot2} and the SNR is significantly
improved.

\begin{figure*}
\centerline{\includegraphics[width=.7\linewidth]{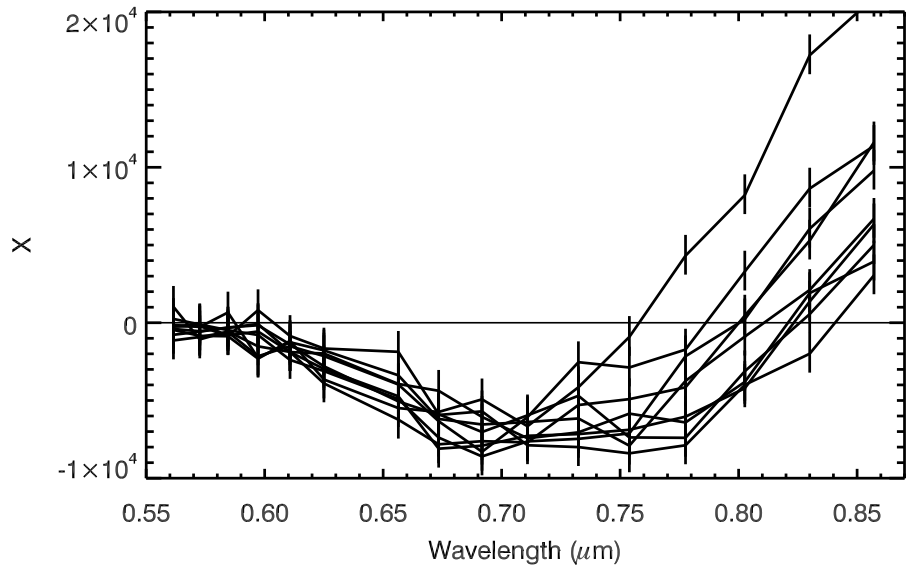}}
\caption{\label{figure_all_scans_real} Plot of the real component (X)
  of the visibility of all nine observations. The zero-crossing is clearly
  visible in each observation.}
\end{figure*}

\begin{figure*}
\centerline{\includegraphics[width=.7\linewidth]{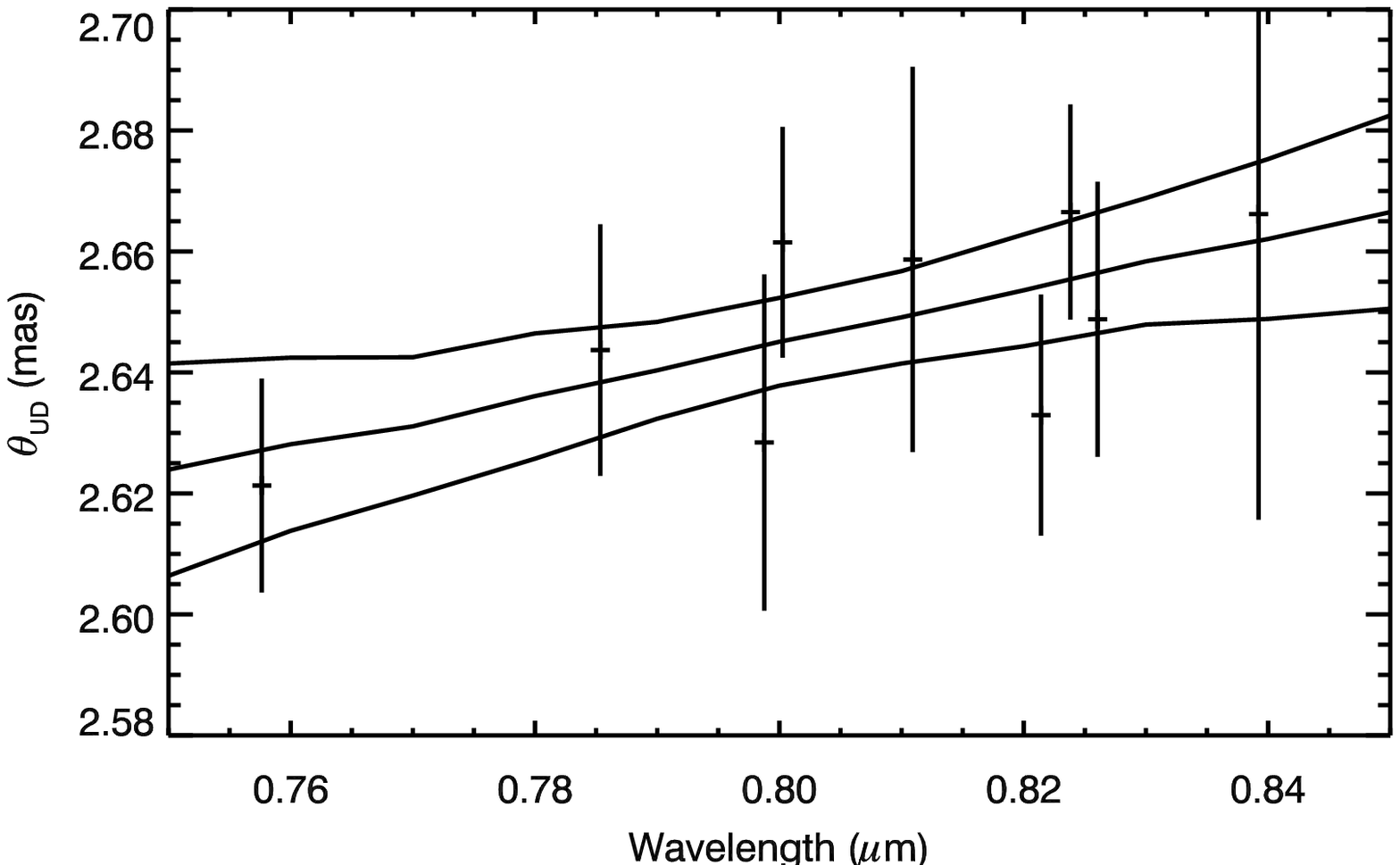}}
\caption{\label{figure_diameter_vs_wavelength} Diameter of the star as
  a function of wavelength (solid curve) and the uncertainty range out
  to one standard deviation (dotted curves). At $\lambda=0.80\,\mu\text{m}$
  the diameter is determined to better than one part in 350.}
\end{figure*}

We can repeat this procedure for all nine observations and obtain a
visibility with good SNR on nice observations of the long
baseline. Because the nine observations occur at different hour angle,
the projected baseline length will be different. Thus, we expect the
zero-crossings to occur at different wavelengths, which we do in-fact
see in Figure~\ref{figure_all_scans_real}. Measuring the location of
the zero-crossing and converting it to a uniform-disk diameter we can
then plot these as a function of the wavelength of the visibility
zero-crossing, and obtain the star's diameter as a function of
wavelength as illustrated in Figure~\ref{figure_diameter_vs_wavelength}. The
produces a remarkable precision of
$\theta_\text{UD}(0.80\,\mu\text{m})=2.6451\pm0.0073\,\text{mas}$, or a
precision of one part in 364. With only a calibration of the
wavelength scale and no amplitude calibration we have thus determined
the diameter of a star, from nine observations, to a precision of
better than $0.3\%$, and we have also mapped the variation of the
diameter with wavelength.

\section{Measuring binary stars}
\label{section_binary}

\noindent Binary stars contain a visibility phase signature in
addition to the amplitude signature. It is possible to extract the
binary star parameters while considering only the visibility
phase. The advantage of using the visibility phase is that it does not
suffer from the biases of the visibility amplitude such that the phase
often is much more accurate. A separate paper in these proceedings,
Jorgensen et al. (2008)\cite{jorgensen:2008}, discusses the
measurements of binaries using coherently integrated visibility
phases.

\section{Conclusion}

\noindent In this paper we have outlined the current approach to
coherent integration at the NPOI. The meaning of the visibility phase
is now well-understood, the visibility amplitudes can be calibrated,
and the resulting coherently integrated visibilities can be used to
make high-precision scientific measurements, as illustrated by the
measurement of the diameter - and it's variation with wavelength - of
$\nu\,\text{Ophiuchus}$.

\acknowledgments

The NPOI is funded by the Office of Naval Research and the
Oceanographer of the Navy. This work was also supported by New Mexico
Institute of Mining and Technology.

\bibliographystyle{spiebib}
\bibliography{main}

\end{document}